\documentclass[fleqn,usenatbib]{mnras}

\usepackage{newtxtext,newtxmath}

\usepackage[T1]{fontenc}
\usepackage{ae,aecompl}

\usepackage{graphicx}	
\usepackage{amsmath}	
\usepackage{amssymb}	
\usepackage{xcolor}
\usepackage{comment}
\definecolor{seagreen}{rgb}{0.190, 0.525, 0.361}
\definecolor{darksalmon}{rgb}{0.914, 0.588, 0.478}
\definecolor{steelblue}{rgb}{0.274 0.510 0.706}
\definecolor{purple}{rgb}{0.5 0.0 0.5}
\definecolor{darkred}{rgb}{0.8 0.0 0.0}
\definecolor{darkgold}{rgb}{0.8 0.543 0.0}
\definecolor{disagreeablegray}{rgb}{0.10 0.12 0.14}
\newcommand{\MP}[1]{\textcolor{seagreen!100}{MP: #1}}

\newcommand{\REFEREE}[1]{{\textnormal{#1}}}

\title[DL to measure turbulence index]{Measuring the spectral index of turbulent gas with deep learning from projected density maps}

\author[P. Trevisan et al.]{
Piero Trevisan$^{1}$\thanks{E-mail: pierotrevisan.pt@gmail.com},
Mario Pasquato$^{2,3}$,
Alessandro Ballone$^{1,2,3}$,
and  Michela Mapelli$^{1,2,3}$
\\
$^{1}$ Physics and Astronomy Department Galileo Galilei, University of Padova, Vicolo dell'Osservatorio 3, I--35122, Padova, Italy\\
$^{2}$ INAF, Osservatorio Astronomico di Padova, vicolo dell'Osservatorio 5, I--35122 Padova, Italy\\
$^{3}$ INFN- Sezione di Padova, Via Marzolo 8, I--35131 Padova, Italy\\
}

\date{Accepted XXX. Received YYY; in original form ZZZ}

\pubyear{2019}

\begin{document}
\label{firstpage}
\pagerange{\pageref{firstpage}--\pageref{lastpage}}
\maketitle

\begin{abstract}
Turbulence plays a key role in star formation in molecular clouds, affecting star cluster primordial properties. As modelling present-day objects hinges on our understanding of their initial conditions, better constraints on turbulence can result in windfalls in Galactic archaeology, star cluster dynamics and star formation. Observationally, constraining the spectral index of turbulent gas usually involves computing spectra from velocity maps. 
Here we suggest that information on the spectral index might be directly inferred from \REFEREE{column density maps (possibly obtained by dust emission/absorption)} through deep learning. We generate mock density maps from a large set of adaptive mesh refinement turbulent gas simulations using the hydro-simulation code RAMSES. We train a convolutional neural network (CNN) on the resulting images to predict the turbulence index, optimize hyper-parameters in validation and test on a holdout set. Our adopted CNN model achieves a mean squared error of $0.024$ in its predictions on our holdout set, over underlying spectral indexes ranging from $3$ to $4.5$. We also perform \REFEREE{robustness tests by applying our model to altered holdout set images, and to images obtained by running simulations at different resolutions}.
This preliminary result on simulated density maps encourages further developments on real data, where observational biases and \REFEREE{other} issues need to be taken into account. 
\end{abstract}

\begin{keywords}
methods: statistical -- methods: numerical -- turbulence -- hydrodynamics
\end{keywords}



\section{Introduction}
Machine learning is finding ever new applications in astronomy, ranging from exoplanets \citep[e.g.][]{davies2015oscillation} and variable stars \citep[e.g.][]{armstrong2015k2} to cosmic ray propagation \citep[e.g.][]{johannesson2016bayesian}, the chaotic three-body problem \citep[][]{2019arXiv191007291B}, \REFEREE{star clusters \citep[][]{2016A&A...589A..95P, pang2020different}} and  black holes \citep[][]{2019MNRAS.485.5345A}. The subfield of Deep Learning (DL) has recently found successful application to astronomical problems involving images using convolutional neural networks \citep[CNN;][]{fukushima1982neocognitron, lecun1989backpropagation, lecun1998gradient}, for example in detecting gravitational lenses \citep[][]{2017Natur.548..555H}.
Turbulence has a strong impact on the properties of all phases of the interstellar medium \citep[e.g. see the reviews by][]{Elmegreen04,Scalo04,Hennebelle12}, likely playing a fundamental role in regulating star formation in molecular clouds \citep[][]{MacLow04,Krumholz05,Ballesteros-Paredes07, Federrath12, Hopkins13, Semenov16, Burkhart18}, \REFEREE{and also affecting cosmic ray propagation \citep[see][and references therein]{2015ARA&A..53..199G}, accretion disc physics \citep[][]{1973A&A....24..337S} and the intergalactic medium \citep[see e.g.][]{2011MNRAS.413.2721E, 2011MNRAS.414.2297I}}.
\citet{Koch19} review several techniques used to constrain turbulence properties in observations. The most widespread ones are based on the analysis of power-spectra and correlation of the density and/or velocity \citep[e.g.][]{Scalo84,Stanimirovic99,Lazarian00, Esquivel05, Padoan06,Burkhart09,Chepurnov10}, wavelet decomposition of the density/velocity field \citep[e.g.][]{Gill90,Stutzki98,Ossenkopf01,Ossenkopf08}, probability distribution functions of the density \citep[e.g.][]{Vazquez-Semadeni94,Miesch95,Vazquez-Semadeni97,Ostriker01,Federrath08,Kainulainen11,Schneider15}, or principal component analysis of the density+velocity \citep[e.g.][]{Brunt02a,Brunt02b,Roman-Duval11}.
Recently \citet{2019arXiv190500918P} have shown that a convolutional neural network can distinguish between different levels of magnetization in density maps of turbulent magnetized gas. This suggests that mock images representing just density information are already enough to constrain the physics of turbulent gas. In the following, we show that this includes the spectral index of turbulence.

\section{Methods}
\subsection{Set of hydro-simulations}
\label{sec:sims} 
We ran $1000$ simulations with {\sc ramses}\footnote{\url{https://www.ics.uzh.ch/~teyssier/ramses/RAMSES.html}} \citep[][]{2002A&A...385..337T}, an Adaptive Mesh Refinement (AMR) code for self-gravitating magnetized fluid flows. The computational domain consists of a $10\times{}10\times{}10$~pc box with periodic boundaries, completely filled with uniform density gas ($6.77 \times 10^{-22}$ g/cm$^3$; for a total mass of $10^4 M_{\odot}$). The gas was kept isothermal at $T = 10$~K throughout the simulation. At the beginning of the simulation, we injected a divergence free, mildly supersonic (Mach number $M = \sqrt{2}$) velocity field with power-spectrum index $n$ extracted uniformly between $3.0$ and $4.5$. \REFEREE{This range of spectra includes both the index predicted by Kolmogorov ($11/3$) and Burgers ($4.0$) turbulence. We chose to extend the range further out rather than limiting it between these two values to increase the variability of the training set and also in consideration of turbulence models that predict very different values of the spectral index, such as Iroshnikov-Kraichnan turbulence \citep[][]{1963AZh....40..742I, 1966PhFl....9.1728K}}. Then, we let the system evolve for $0.5$~Myr, solving Euler's equation with a Lax-Friedrichs Riemann Solver, \REFEREE{without self-gravity or magnetic fields}. The AMR strategy in {\sc ramses} allowed us to perform our simulations with a relatively low number of cells (compared to a uniform grid), leading to an affordable computational cost for the whole large set of simulations needed for training. AMR might not seem the best choice to study turbulence, since the smallest scales will not be resolved throughout the whole computational domain. However, a ``smart'' refinement criterion allows to have high resolution on the regions of the computational domain that are more physically meaningful. In these simulations we were interested in how the velocity field shapes the density by means of gas collision. Hence, we chose the refinement criteria based on the gradient of the velocity: for each cell $i$, the gradient of velocity $v$ is computed using the six nearest-neighbouring cells. If this gradient, times the local mesh spacing $\Delta x^l$ at level of refinement $l$, exceeds a fraction of the central cell variable: 
\begin{equation}
    \nabla v_i \ge C_v \frac{v_i}{\Delta x^l}, 
\end{equation}
where $C_v$ is a free parameter, then the cell is refined to level $l+1$ \citep{2002A&A...385..337T}. For our set of simulations, we chose $C_v = 1.35$. We adopted this fixed value for all simulations, after testing that this choice allowed to effectively resolve  the turbulence for all values of $n$ with a reasonably high number (always $>10^5$) of cells. We set the minimum and maximum refinement levels as 5 and 8, respectively. This meant a spatial resolution of $2^{-5} = 1/32$ of the box side ($\approx$ 0.3 pc) for the least resolved cells and a resolution of  $2^{-8} = 1/256$ of the box side ($\approx$ 0.04 pc) for the most resolved ones.

\subsection{Mock image generation}
\begin{figure}[h!]
    \centering
    \includegraphics[width=0.95\columnwidth]{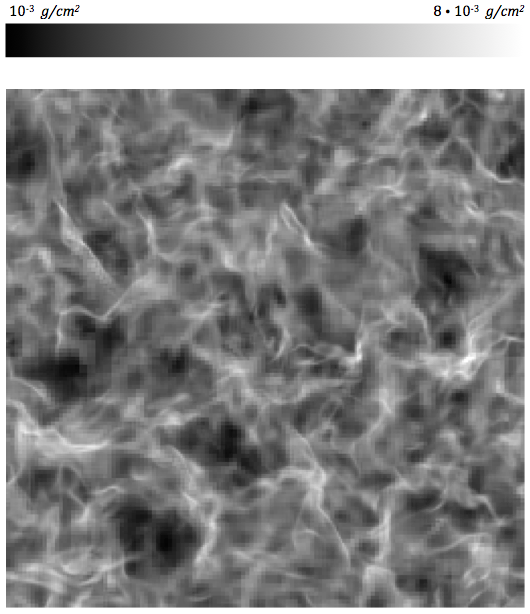}
    \caption{Projected density map along the $z$ axis for the simulation with the lowest $n = 3.0013$. Pixel intensity values are normalized to 1. The density to pixel intensity transformation is logarithmic. The generated image is $1000\times{}1000$ pixels in size, in one channel (grayscale).}
    \label{maxn}
\end{figure}

\begin{figure}
    \centering
    \includegraphics[width=0.95\columnwidth]{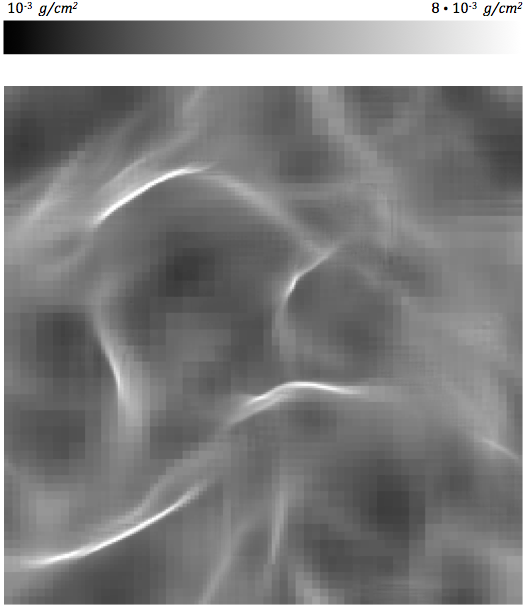}
    \caption{Projected density map along the $z$ axis for the simulation with the highest $n = 4.4997$. The color map and resolution is the same as in Figure~\ref{maxn}.}
    \label{minn}
\end{figure}

For each simulation, we took snapshots of the column density projected onto three perpendicular axes of the computational domain, allowing us to increase the number of images obtained from each simulation. However, we took care that each set of three resulting images with the same index ended up together either into the training set or into the holdout set, so that our models are tested not just on previously unseen \emph{images}, but on unseen \emph{simulations}. \REFEREE{Images were generated from a given snapshot by integrating through the entire density cube along three orthogonal directions resulting in column density maps (as could be obtained e.g. by dust emission/absorption). This is a first step towards the use of machine learning models for future, more sophisticated analyses, taking into further account the velocity information (coming, in real observations, from molecular tracers). 
Integrated column density maps represent a harder problem for our CNNs because integration strongly reduces the imprint of the velocity field by removing velocity information \citep[see e.g.][]{2000ApJ...537..720L, 2004ApJ...616..943L, 2006ApJ...652.1348L, 2008ApJ...686..350L, 2018ApJ...865...46L}.}
\REFEREE{To illustrate the results of our image generation procedure,} figure~\ref{maxn} and ~\ref{minn} show two images with low and high $n$, respectively.

\subsection{Deep learning: neural network architecture}
Since the deep convolutional neural network AlexNet \citep[][]{krizhevsky2012imagenet} won the 2012 ImageNet competition \citep[][]{ILSVRC15}, CNNs have been successfully applied to a large variety of computer vision tasks, such as e.g. object detection \citep[][]{girshick2014rich}, where they routinely outperform other machine learning approaches \citep[see e.g. ][for more details]{Goodfellow-et-al-2016}. 

We implemented our neural networks in Keras \citep[][]{Charles2013}, on top of TensorFlow \citep{tensorflow2015-whitepaper}. All our code is written in Python and can be found at \textsf{https://gitlab.com/Piero3/turbolencenn}.

The neural network architecture we chose is as follows:
\begin{itemize}
    \item Two $5\times{}5$ convolutional layers with 32 filters and \emph{same} padding resulting in a $128\times 128\times 32$ output; 
    \item A max pooling layer with $2\times 2$ filter size and stride 2, resulting in an $64\times 64\times 32$ output;
    \item A dropout layer with a variable ratio of dropped units (either $0$ or $1/3$).
    \item Two $5\times5$ convolutional layers with $64$ filters and same padding resulting in a $64\times 64\times 64$ output;
    \item A max pooling layer with $2\times2$ filter size and stride $2$, resulting in an $32\times 32\times 64$ output;
    \item A dropout layer with a variable ratio of dropped units (either $0$ or $1/3$).
    \item A fully connected (dense) layer with $64$ neurons with rectified linear unit activation.
    \item A dropout layer with a variable ratio of dropped units (either $0$ or $1/3$).
    \item A single neuron with a linear activation with relative squared error loss as cost function.
\end{itemize}

This architecture was chosen after some trial-and-error experimentation starting from a very shallow initial configuration and progressively adding more layers. Aftwerwards we performed a grid search over the few hyperparameters that we decided to explicitly optimize, as discussed below.

\subsection{Deep learning: hyperparameter optimization}
We train our neural networks on $896$ simulations corresponding each to a different spectral turbulence index. Each simulation is projected on three independent directions, obtaining three column density map images. The remaining $104$ simulations are set aside to perform a blind test of the accuracy of the trained networks; these simulations were also never used during the initial tests we conducted to determine the overall network architecture. We optimized the hyperparameters of our nets using an $80\%-20\%$ train-validation random split.
We trained our network for different choices of the dropout ratio: either $1/3$ for all dropout layers or zero (corresponding to no dropout), and we compared four optimizers: AdaDelta  \citep[][]{zeiler2012adadelta}, AdaGrad \citep[][]{duchi2011adaptive}, RMSprop \citep[][]{tieleman2012lecture} and Adam \citep[][]{kingma2014adam}. We train our CNNs for either $500$ epochs (if the dropout is set to $1/3$) or for $100$ epochs (if trained with no dropout). This is done to reduce overfitting for models without dropout in a sort of early stopping scheme \citep[e.g. see ][]{prechelt1998automatic}. We also consider different batch sizes: ($32$, $64$, $128$, $256$). We show the resulting training and validation Mean Squared Error (MSE) loss in Tab.~\ref{hyper}, where each training run is sorted by increasing validation MSE. In this phase we used an $80-20~\%$ train-validation split.

\begin{figure}
    \centering
    \includegraphics[width=0.9\columnwidth]{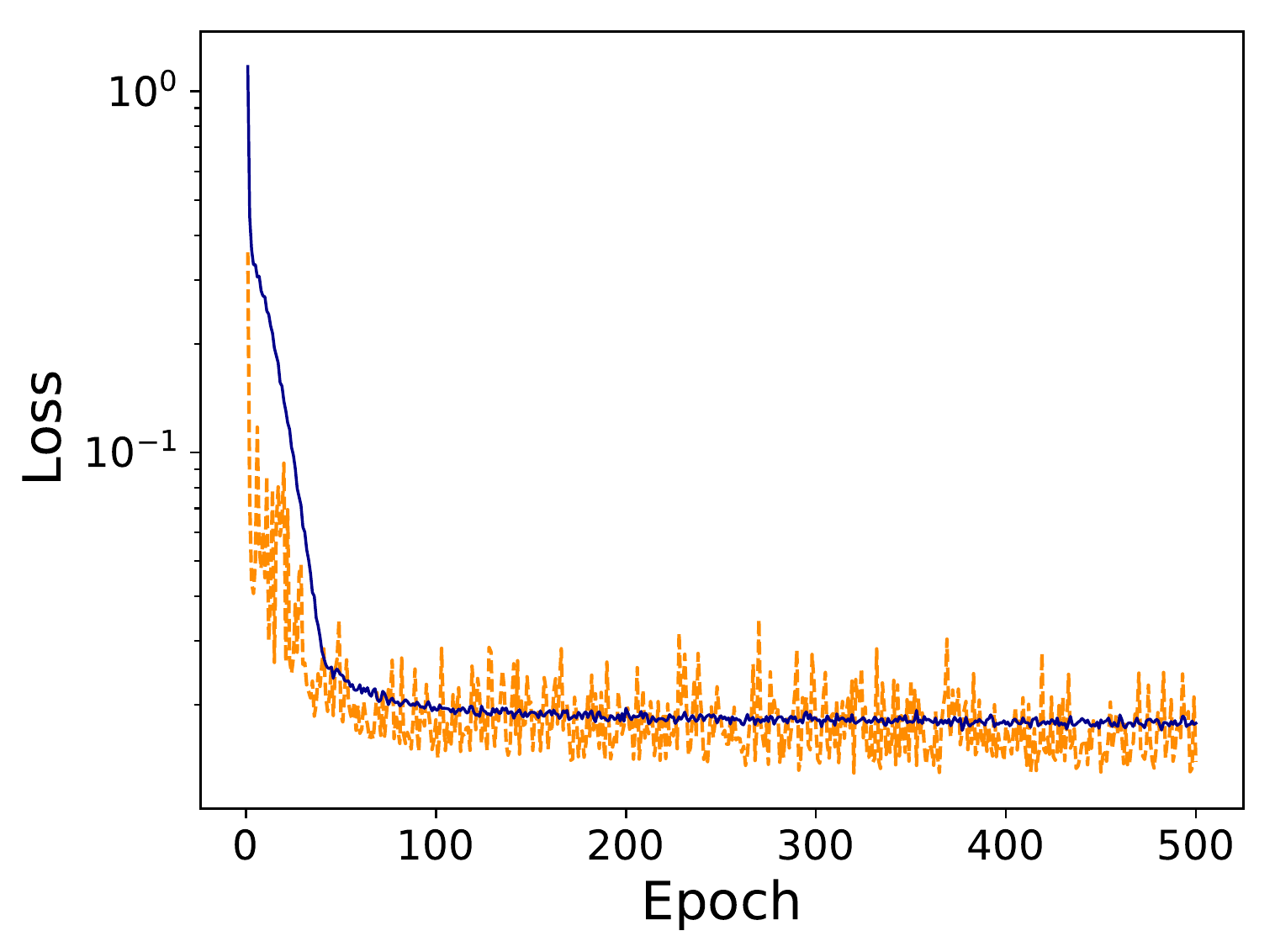}
    \caption{Evolution of the training loss (solid blue curve) and validation loss (dashed orange curve) during training for the adopted CNN.}
    \label{fig:loss1000withzoom}
\end{figure}

Our best performing model (in terms of validation loss) is the result of training with the Adadelta optimizer \citep[][]{2014arXiv1412.6980K} for $500$ epochs with a batch size of $64$ and $1/3$ dropout. The validation MSE loss at the end of training is $1.2 \times 10^{-2}$. However, this model is clearly overfitting, as the validation loss is much higher than the training loss. The best model that is not overfitting (as shown by it having a higher training loss than validation loss) is the third best in terms of validation loss, achieving a final MSE of $1.4 \times 10^{-2}$, and is the result of training with the Adam optimizer for $500$ epochs with a batch size of $64$ and $1/3$ dropout. We adopt this latter model and use it in the following. The evolution of the training and validation loss for this model is shown in Figure \ref{fig:loss1000withzoom}.

At this point our CNN never saw our holdout set of $312$ column density snapshots (corresponding to $104$ different indexes). This also holds true for our previous, informal optimization that yielded the CNN architecture we adopted in the first place.

\begin{table}
\centering
\caption{Summary of our hyperparameter optimization: each row corresponds to a different combination of hyperparameters. The MSE loss after either $500$ training epochs (models trained with dropout) or $100$ epochs (models without dropout) is reported for the training set and for the validation set (last two columns) as a function of the dropout fraction (first column), the optimizer (second column) and the batch size used in training (third column).\label{hyper}}
\begin{tabular}{llllll}
Dropout & Optimizer & Batchsize & Epochs & Train & Val. \\
        &           &           &        &loss      &loss \\
\hline
0.33  & adadelta & 64 & 500 & 0.005 & 0.012\\
0.33  & adadelta & 32 & 500 & 0.006 & 0.014\\
0.33  & adam & 64 & 500 & 0.018 & 0.014\\
0.33  & adam & 256 & 500 & 0.014 & 0.014\\
0.00  & adam & 32 & 100 & 0.001 & 0.014\\
0.33  & rmsprop & 256 & 500 & 0.015 & 0.016\\
0.33  & rmsprop & 32 & 500 & 0.021 & 0.016\\
0.00  & adam & 128 & 100 & 0.002 & 0.016\\
0.00  & adam & 64 & 100 & 0.001 & 0.016\\
0.33  & rmsprop & 128 & 500 & 0.020 & 0.017\\
0.00  & rmsprop & 32 & 100 & 0.005 & 0.018\\
0.00  & adadelta & 32 & 100 & 0.005 & 0.021\\
0.00  & rmsprop & 64 & 100 & 0.013 & 0.021\\
0.00  & adadelta & 64 & 100 & 0.010 & 0.022\\
0.00  & adagrad & 64 & 100 & 0.008 & 0.023\\
0.33  & adam & 128 & 500 & 0.014 & 0.024\\
0.33  & adam & 32 & 500 & 0.022 & 0.024\\
0.00  & adagrad & 32 & 100 & 0.002 & 0.027\\
0.00  & adagrad & 256 & 100 & 0.027 & 0.034\\
0.00  & adagrad & 128 & 100 & 0.022 & 0.039\\
0.00  & adadelta & 128 & 100 & 0.035 & 0.047\\
0.00  & rmsprop & 128 & 100 & 0.018 & 0.052\\
0.00  & rmsprop & 256 & 100 & 0.025 & 0.063\\
0.00  & adadelta & 256 & 100 & 0.095 & 0.085\\
0.33  & adadelta & 128 & 500 & 0.190 & 0.190\\
0.33  & adadelta & 256 & 500 & 0.190 & 0.190\\
0.33  & rmsprop & 64 & 500 & 0.190 & 0.190\\
0.33  & adagrad & 128 & 500 & 0.242 & 0.470\\
0.33  & adagrad & 32 & 500 & 0.535 & 0.534\\
0.33  & adagrad & 64 & 500 & 0.284 & 0.763\\
0.33  & adagrad & 256 & 500 & 0.295 & 0.980\\
0.00  & adam & 256 & 100 & 3.622 & 3.511\\
\end{tabular}
\end{table}

\subsection{Deep learning: data augmentation}\label{augm}
Since running hydrodynamic simulations can be very time consuming and computationally expensive, we augmented the training dataset by applying transformations such as cropping, reflections and rotations. To perform the data augmentation we used the python library \textsf{Augmentor}\footnote{\url{https://github.com/mdbloice/Augmentor}}. With this process we artificially enlarged our dataset from $2688$ to $20000$ images. 
The number of combinations of cropping plus reflections and rotations guarantees that we do not have two identical examples in our training dataset or across the training/validation split. The holdout set, which was used neither in training nor in validation, did not undergo augmentation.

\section{Results}
\begin{figure}
    \centering
    \includegraphics[width=0.9\columnwidth]{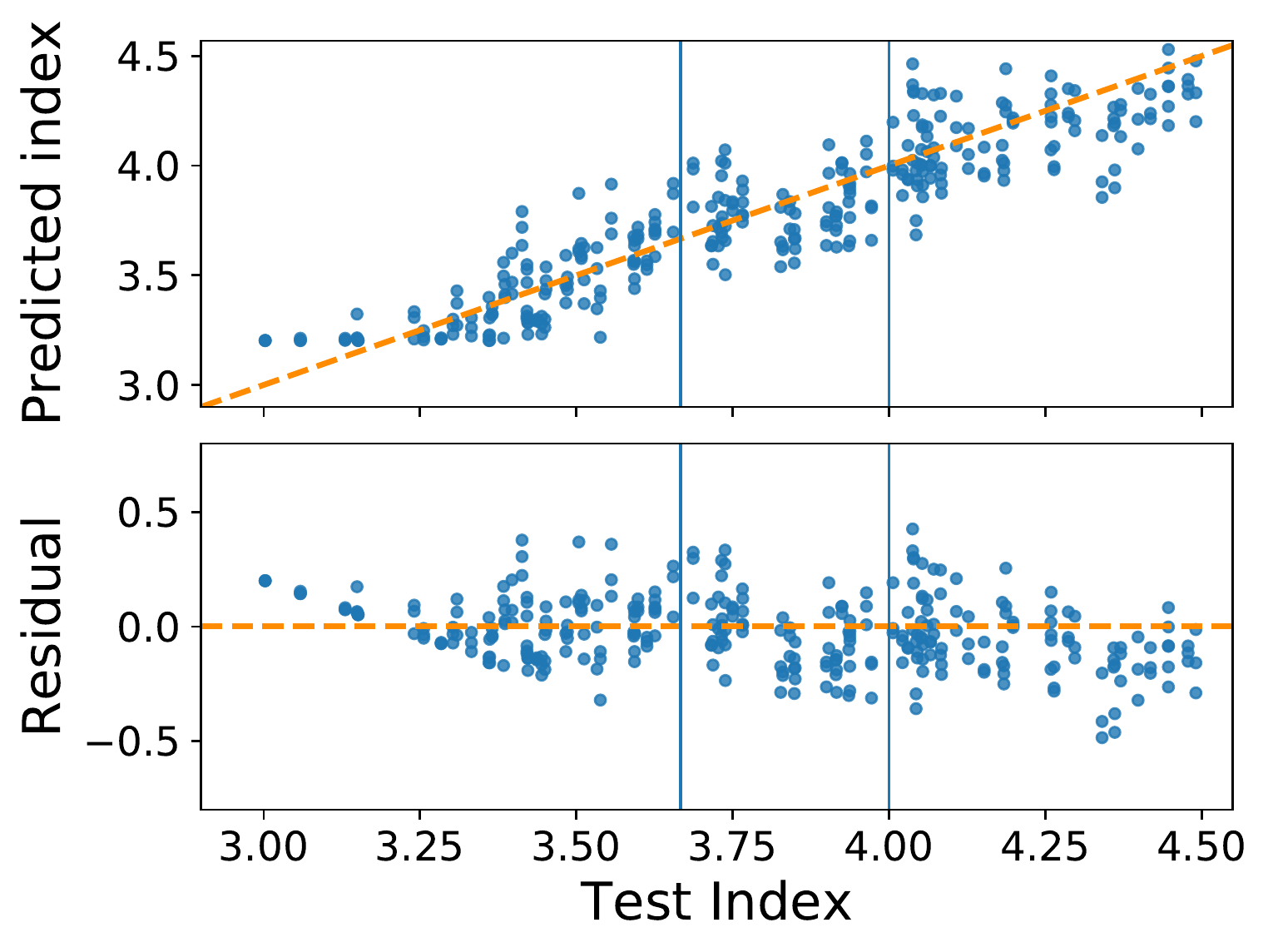}
    \caption{Prediction of the adopted CNN model on the $312$ test images in the holdout set. The CNN was not shown these images in training nor in validation, neither was it shown images derived from the same simulations as these. Top panel: power spectrum indexes predicted from the CNN plotted versus the actual indexes labeled as test indexes. Bottom panel: the residual are plotted  versus the test indexes. The vertical line at $n=11/3$ and $n=4.0$ corresponds to Kolmogorov and Burgers index respectively.}
    \label{fig:scatterplot}
\end{figure}

We re-trained our adopted CNN (trained with $1/3$ dropout fraction, the Adam optimizer, batch size of $64$; see third row of Tab.~\ref{hyper}) for $500$ epochs on the whole new training dataset, described in section \ref{augm}.
After training, we tested our CNN on our holdout set of $312$ images resulting from simulations made with the same ingredients of the training simulations.
The $312$ images correspond to $104$ simulations with different turbulence index, seen from three different perpendicular directions. On this set, we did not perform any augmentation process (crop, flip, rotation) as opposed to what we did in training. The predictions of our adopted CNN are shown in Fig.~\ref{fig:scatterplot}. We obtained a mean-squared error (MSE) between our predictions and the actual spectral turbulence indexes of $0.024$. 

\subsection{Robustness tests: \REFEREE{low-pass filter}}
\begin{figure}
    \centering
    \includegraphics[width=0.95\columnwidth]{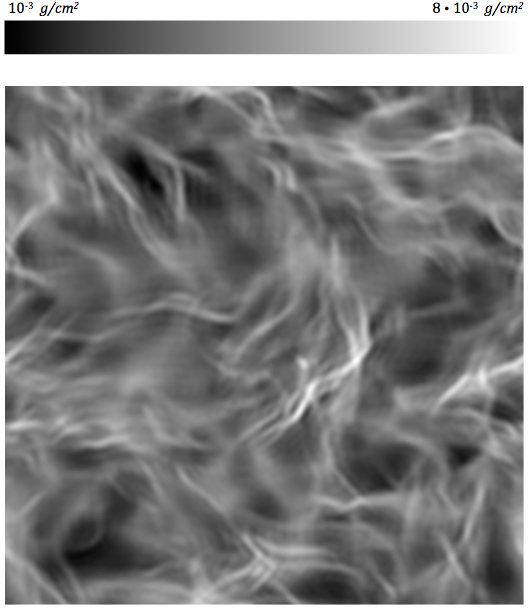}
    \caption{Typical input for our low-pass filter experiment. Spatial frequencies above a threshold are cut off in Fourier space. Different cutoffs were experimented with.}
    \label{fig:lo}
\end{figure}

While our model performs well in the idealized setting we considered, it is expected that regression accuracy will drop in realistic conditions, e.g. when attempting to predict the spectral index of turbulent gas from actual observations. As a first test of robustness under less than ideal conditions we degrade the density maps in our holdout set and measure the resulting drop in performance of our model. 
We apply Fourier transform to our images and remove high spatial frequencies (low-pass filter) at different cutoffs. A low-pass filter blurs out the fine spatial structure of the density map, which has a similar effect to reducing the resolution of the image or convolving with an instrumental point spread function, as shown in Fig.~\ref{fig:lo}. Mesh artifacts (e.g. sharp discontinuities in density at cell boundaries) are largely removed by low-pass filtering, as the typical cell size is on the small end of the spatial frequency range for our images. 

\begin{figure}
\centering
    \includegraphics[width=0.95\columnwidth]{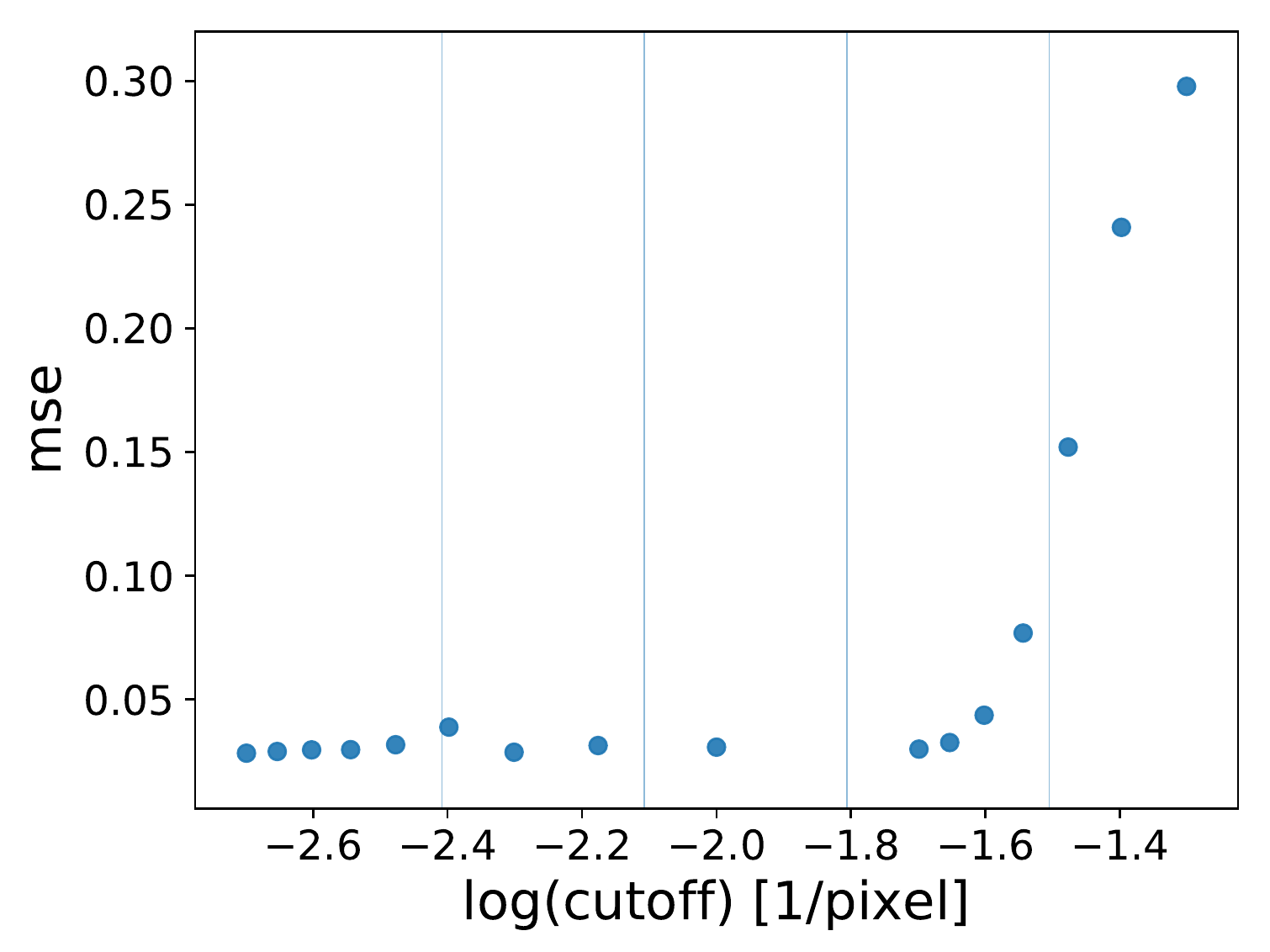}
    \caption{MSE as a function of the spatial frequency cut-off (in $1/$pixels) of the low-pass filter, i.e. frequencies below the cut-off are kept, so the leftmost point corresponds to retaining the original image.}
    \label{msefilter}
\end{figure}
Fig.~\ref{msefilter} shows the MSE of our adopted model on the test set low-passed at various cutoff spatial frequencies. The rightmost point corresponds to the unaltered images. As we move left and the cutoff is lowered, MSE increases (performance drops) but not abruptly so. In particular, the relevant mesh frequencies do not stand out as discontinuity points in this graph. This suggests that the model is not picking up on high-frequency mesh artifacts to make its predictions.

\subsection{\REFEREE{Robustness tests: resolution convergence}}
\begin{figure}
\centering
    \includegraphics[width=0.95\columnwidth]{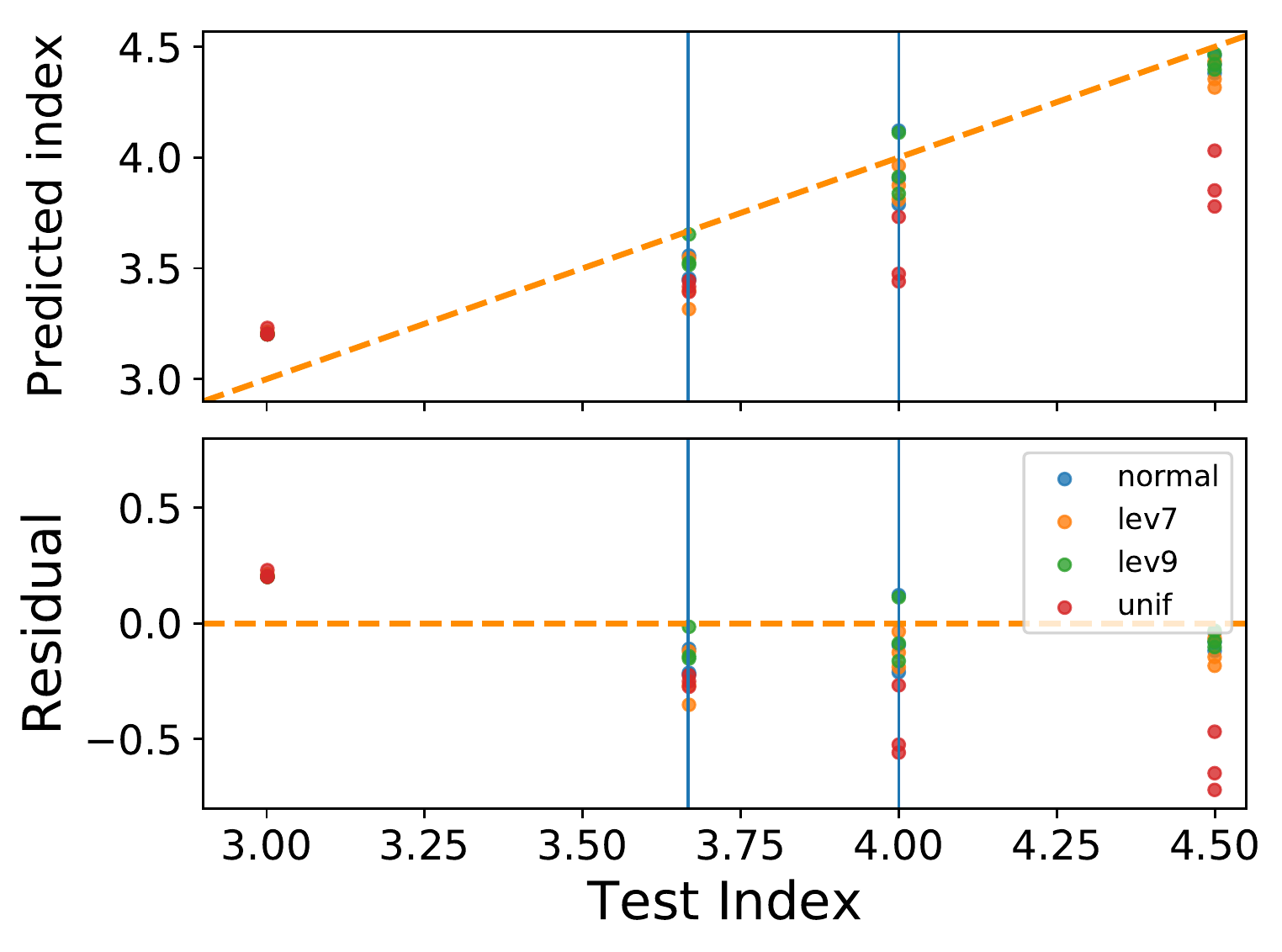}
    \caption{\REFEREE{Predicted spectral indexes as a function of the actual indexes for three projections of the four simulations we 
    reran with different resolution. The original simulations are shown as blue dots, AMR simulations with lower (higher) resolution limit in yellow (green), and uniform grid simulations in red. The diagonal line is the identity.}}
    \label{resimulate}
\end{figure}

\REFEREE{Perhaps a more stringent requirement than performing well on images with a different resolution would be to make correct predictions on \emph{simulations} run at a different resolution. Due to the computational costs of hydrodynamic simulations, we selected only four simulations corresponding to indexes near to the range extremes $3.0$ and $4.5$ and to the Kolmogorov ($11/3$) and Burgers regime ($4.0$). We re-ran these simulations with different AMR resolution limits, namely $2^{-7}$ the box side (thus lowering the maximum attainable resolution with respect to the original) and $2^{-9}$ the box side (increasing the maximum resolution). We also run a simulation with a fixed, uniform mesh with cells $2^{-8}$ the box side}. \REFEREE{The predictions of our best CNN model, which is trained only on simulations run with AMR and maximum resolution corresponding to $2^{-8}$ the box side, are shown in Fig.~\ref{resimulate} as a function of the actual spectral indexes. As usual, we calculated our predictions on three independent projections of each simulation, so each index corresponds to three points in the plot. We see that prediction accuracy is as high as on the original simulations for the new, higher- and lower-resolution AMR simulations, while it drops somewhat for the uniform mesh. This simple qualitative test suggests that the model predictions are robust with respect to changes in resolution, even though making a quantitative statement in this regard would require rerunning a larger sample of simulations.}

\section{Discussion and conclusions}
We have trained a convolutional neural network to predict the spectral index of turbulence of mock column density maps generated by simulations of turbulent gas. Our neural network model accomplishes this task by using only pixel-level information from images. 

With a fixed five-layer feedforward network architecture, we obtain a performance of $0.024$ in terms of mean squared error on an holdout test set unseen in training, spanning spectral indexes from $n=3$ to $n=4.5$. This is an encouraging result as it suggests that plain density maps contain sufficient information for an accurate prediction of the spectral index of turbulence. \REFEREE{Moreover, our images are generated by fully projecting the density distribution of the gas along the line of sight, essentially disregarding velocity information.} \REFEREE{With this in mind, even though the mean squared error we obtain is still at face value too high for observational applications (Kolmogorov and Burgers indexes differ only at the two-sigma level), we expect to reduce it by averaging predictions obtained on independent regions of a given cloud.}

To ascertain that our model is indeed using relevant physical information to obtain its predictions we ran a series of tests by degrading our test images by censoring high spatial frequencies and measuring the resulting drop in performance. We find that blurring out the fine spatial structure of our images (including any mesh artifacts such as abrupt changes in density at projected cell boundaries) in this way progressively lowers our model's performance, but we do not observe sharp jumps at mesh frequencies, suggesting that the model is not using simulation artifacts to drive its predictions. \REFEREE{Additionally, we re-run a handful of simulations with different mesh resolutions, obtaining accurate predictions on the derived images, further supporting the robustness of our results.} \REFEREE{Possible future developments of this work along these lines are related to using machine learning interpretability techniques on our trained model to reveal explanations as to why a given prediction is cast: intelligible explanation are as important as accuracy in scientific applications.}

\REFEREE{While these checks suggest that the model is not picking up subtle clues from simulation artifacts, there are still several issues that we need to address before applying this model to actual data: first of all, we need to first identify which observational data are more suitable to be adopted for this analysis. For example, previous theory and numerical studies have shown that in the case of optically thick tracers the spectral index saturates to $-3$ \citep[][]{2004ApJ...616..943L, Burkhart_2013}, so our CNN might never be able to predict either the density or velocity spectral index. Moreover,} the simulations we considered are highly idealized, lacking \REFEREE{important} physical ingredients such as magnetic fields, \REFEREE{relevant chemical reaction networks}, and self-gravity. For this proof-of-concept work, we justify this choice based on the much higher computational resources needed to model these ingredients. However, the entity of the bias affecting a deep learning model trained on simplified simulations when applied to real data is still in need of quantification: a first check could be to run a limited number of more physically realistic simulations and evaluate the accuracy of our model predictions on them. Irrespective of the sophistication of their physics, another limitation of simulations is their resolution, which even with AMR cannot fully cover the range of scales spanned by turbulent gas in real systems. However, these issues are shared with any modelling that relies on simulations and are not directly related to our machine learning approach, \REFEREE{which incidentally yields accurate results even on simulations run with a different resolution with respect to the one used in training.}

Our CNN approach has different strengths and weaknesses, as opposed to more time-tested approaches such as directly fitting the density power spectrum (obtained e.g. by fast-Fourier transforming an image). For example the latter method, while simpler and easier to interpret, requires some discretion in determining the power-law region of the spectrum to consider, e.g. by setting fiduciary cutoffs in spatial frequencies above and below which the spectrum data is disregarded. Additionally, our networks can be easily repurposed to predicting different physical quantities of the turbulent gas which may not be immediately accessible to spectral methods.


\section*{Acknowledgements}
This project has received funding from the European Union's Horizon $2020$ research and innovation programme under the Marie Sk\l{}odowska-Curie grant agreement No. $664931$. AB, and MM acknowledge financial support by the European Research Council for the ERC Consolidator grant DEMOBLACK, under contract no. 770017. MP wishes to thank Prof. David W. Hogg, Dr. Gregor Seidel and Prof. Stella Offner for feedback and discussion.

\section*{Data availability}
The data underlying this article will be shared on reasonable request to the corresponding author.
\bibliographystyle{mnras}
\bibliography{MLturbulence} 

\bsp	
\label{lastpage}
\end{document}